\documentstyle[12pt]{article}
\newcommand{\be}{\begin{equation}}
\newcommand{\ee}{\end{equation}}
\newcommand{\bn}{\begin{eqnarray}}
\newcommand{\en}{\end{eqnarray}}
\newcommand{\bd}{\begin{displaymath}}
\newcommand{\ed}{\end{displaymath}}
\begin{document}
\title{Self - Interaction Force for the Particle in the Cone
Space - Time}
\author{N R Khusnutdinov\thanks{e-mail:nail@univex.kazan.su}\\\\
{\em Department of Theoretical Physics,}\\{\em Kazan State
Pedagogical Institute, Kazan,}\\{\em P.O.Box 21, Mezhlauk St., 1,
420021, Russia}}
\date{}
\maketitle
\begin{abstract}The force acting on the charged particle moving along
an arbitrary trajectory near the straight cosmic string is
calculated.  This interaction leads to the scattering of particles by
the cosmic string. The scattering cross section is considered.
\end{abstract}
\section{Introduction}
Cosmic strings may have resulted from phase transitions in the early
Universe [1,2]. Stability of these formations is insured by the
appropriate non-zero topological charge (winding number).  From the
well-known solutions of field equations describing gravitating cosmic
strings [3-7] one can see that the energy density is located in a
small threadlike region of space, which is why one may approximate
them by infinitely thin curves. In this case the cosmic strings has
no Newtonian potential [8] and the space-time is flat.

At the same time the non-local display of cosmic strings exists. The
cosmic string produces double images through gravitational lensing [8
-10]. If a resting particle has electric charge then there will be a
repulsive force between the string and the particle [11].  A charged
particle radiates electromagnetic waves even if it moves uniformly
near the cosmic string [12]. There is also the Aharonov-Bohm
interaction of cosmic strings with matter [13,19].

In this article we analyse in more detail the force of self -
interaction of a charged particle near a straight cosmic string. We
obtain this force for the arbitrary trajectory of the particle. In
the simplest case of a particle at rest we recieve a result which
differs from that obtained in Ref.11.
\section{Solution of Maxwell's equations}
The spacetime of a straight cosmic string is described by the metric
[8]
\be
ds^{2} = dt^{2}-dz^{2}-d\rho^{2}-b^{2}\rho^{2}d\varphi^{2}\ ;\ b \leq
1\ .
\ee
Space\-time with met\-ric (1) is the pro\-duct of two di\-men\-sional
pseudo - Euclidian subspace $(t,z)$ and two dimensional Euclidian
cone space $(\rho,\varphi)$, which allows us the Maxwell's equations
\be
F^{ik}_{;k} = -\frac{4\pi}{c}J^{i} = -\frac{4\pi e}{\sqrt{-g}}\int\!
d\tau\,\delta^{(4)}\,(x-x(\tau))\,u^{i}\,(\tau)\ ;\ F_{(ik;l)} = 0 \
,
\ee
in the Lorentz gauge split into two systems of equations
\be
(\partial^{2}_{t} - \Delta_{(s)})A^{(s)} = \frac{4\pi}{c}J^{(s)}\ ,
\ee
where
\be
\Delta_{(s)} = \frac{1}{\rho}\,\frac{\partial}{\partial\rho}\left
(\rho\frac{\partial}{\partial\rho}\right) + \frac{1}{\rho^{2}}\left
(\frac{\partial}{b\,\partial\varphi} + i\,s\right)^{2} +
\frac{\partial^{2}}{\partial z^{2}}\ ,
\ee
\bd
s = 0,1\ ;\ A^{(1)} = A_{\rho} + \frac{i}{b\rho}A_{\varphi}\ ;\
A^{(0)} = (A_{t}\ ,\ A_{z})\ .
\ed

Taking the Fourier transformation
\bd
A(\omega) = \int^{+\infty}_{-\infty} dt\,A(t)\,e^{-i\omega t}\ ,
\ed
we obtain the following equations
\be
(\Delta_{(s)} + \omega^{2})A^{(s)}({\bf r},\omega) =
\frac{4\pi}{c}J^{(s)}({\bf r},\omega)\ .
\ee

The retarded Green function satisfying the expression
\be
(\Delta_{(s)} + \omega^{2})G^{(ret)}_{(s)}({\bf r},{\bf r}';\omega)
= -\delta(\rho-\rho')\delta(\varphi-\varphi')\delta(z-z')/b\rho\ ,
\ee
has been calculated in Ref.12.

This function has the following form
\bn
G^{(ret)}_{(s)}({\bf r},{\bf r}';\omega)&=&-\int_{-\infty}
^{+\infty}\frac{dk}{2\pi}\,e^{ik\Delta z}\,
\sum_{n=-\infty}^{+\infty}\frac{1}{2\pi b}\,e^{in\Delta
\varphi} \\&\times&\int_{0} ^{+\infty}\frac{\lambda
d\lambda}{(\omega-i\,0)^{2}-k^{2}-\lambda^{2}} J_{|
n\nu+s|}(\lambda\rho) J_{| n\nu+s|}(\lambda\rho')\ .\nonumber
\en
Hereafter we set $\Delta z=z-z',\Delta\varphi =\varphi -
\varphi',\nu =1/b\geq 1$ ; and $J_{p}(x)$ - the Bessel function. The
Green function (7) can be found by expanding over the full set of
eigenfunctions of the operator $\Delta_{(s)}$ .

Then the solution of equation (3) maybe presented in the following
form
\bn
A^{(s)}({\bf r},t) & = & -2e\int_{-\infty}^{+\infty} d\tau
\int_{-\infty}^{+\infty} d\omega \int_{-\infty}^{+\infty}
\frac{dk}{2\pi}e^{i\omega(t-t(\tau))+ik(z-z(\tau))}\nonumber \\
& \times & \sum_{n=-\infty}^{+\infty}\frac{1}{2\pi b}
e^{in\Delta\varphi(\tau)}\int_{0}^{+\infty}
\frac{\lambda   d\lambda}{(\omega-i\,0)^{2}-k^{2}-\lambda^{2}}\\
& \times & J_{| n\nu+s|}(\lambda\rho) J_{|
n\nu+s|}(\lambda\rho(\tau))u^{s}(\tau) , \nonumber
\en
where $u^{(1)}(\tau)=u_{\rho}(\tau)+i\nu u_{\varphi}(\tau)/\rho(\tau)
\ ,\  u^{(0)}(\tau)=(u_{t}(\tau)\ ,\ u_{z}(\tau))$ .

Integrating successively over the $\omega, k, \lambda$ and, where
possible, summing the series we obtain (see appendix) :
\bn
A^{(s)}&=&e\nu\int_{-\infty}^{+\infty} d\tau u^{(s)}(\tau)
\theta (t-t(\tau))\frac{\theta (q)}{\rho\rho (\tau)\sqrt{4q| q-1
|}}\nonumber \\&\times &\left\{ \theta (1-q)Q(s)+\theta
(q-1)Q_{\nu}(s)\right\}\\&=&A_{1}^{(s)} + A_{2}^{(s)}\ ,\nonumber
\en
where
\bd
q = [ (t-t(\tau))^{2}-(z-z(\tau))^{2}-(\rho-\rho (\tau))^{2}] /
4\rho\rho (\tau) , \beta = 2q-1+\sqrt{4q(q-1)} ,
\ed
\bn
Q(s)&=&
\sum_{n=-\infty}^{+\infty}[e^{is\arccos (1-2q)}
\delta (\Delta\varphi (\tau)+2\pi n+\nu\arccos (1-2q))\nonumber \\
&+&
e^{-is\arccos (1-2q)}\delta (\Delta\varphi (\tau)+2\pi n
-\nu\arccos (1-2q))]\ ,\\
Q_{\nu}(s)&=&
-(-1)^{s}\frac{\sin(\pi\nu)}{\pi}\{\cosh(s\ln\beta)[
\cosh(\nu\ln\beta)\cos(\Delta\varphi(\tau))-\cos(\pi\nu)]\nonumber\\
&-&
i\sinh(s\ln\beta)\sinh(\nu\ln\beta)\sin(\Delta\varphi(\tau))\}
/\{[\cosh(\nu\ln\beta)\nonumber \\
&-&
\cos(\Delta\varphi(\tau)-\pi\nu)]
[\cosh(\nu\ln\beta)-\cos(\Delta\varphi (\tau)+\pi\nu)]\}\ .\nonumber
\en

 From (9)-(11) one can see that the electromagnetic field arising from
the particle consists of two parts. The first, singular term,
describes the propagation of radiation along the isotropic geodesics;
this is because the condition of vanishing of the arguments of
$\delta$ - functions is equivalent to zero value of the square of the
geodesic interval between the points of observations and position of
charge.  The sum of $\delta$- functions occurs due to the fact that
in the cone spacetime any two events may be connected by several
geodesic lines.  The smaller the parameter $b$ , the more of these
geodesic lines exist.  Indeed, when $b$ is small, the cone is sharper
and looks like a cylinder locally, but cylindric space has an
infinite number of geodesics connecting two points. If there are
several isotropic geodesics connecting a point of observation and a
position of charge, or if there are several of these, then all of
them must be taken into account.

 From expression (10) one may extract the restrictions on the possible
values of integer $n$. Since the function $\arccos x$ changes from
$0$ to $\pi$, we infer the following expression:
\be
-\nu\pi\leq \Delta\varphi + 2\pi n\leq\nu\pi\ .
\ee
This condition coincides with that from Ref.\,14.

The second non-local term vanishes when $\nu$ is integer. We
emphasize that a particle moving uniformly in spacetime (1) does not
radiate the electromagnetic wave when the analogy condition is
satisfied.  Inequality $q\geq O$ shows that the field at the point of
observation is due to the motion of a particle into the cone of past
events.
\section{Self-interaction force}
In order to obtain this force we use the traditional method with the
help of which the Dirac - Lorentz force has been calculated [15].
Previously, this method has been used to obtain the gravitational -
induced self - interaction force acting on the charged particle
situated in the field of strong plane gravitational wave [16].

At the beginning it is necessary to receive in the integral form the
tensor of electromagnetic field
$F_{ik}=\partial_{i}A_{k}-\partial_{k} A_{i}$ at the point of
observation. The force acting on the charge $e$, moving with velocity
$u^{k}$ at the point of observation has the following form
\be
{\cal F}_{i} = eF_{ik}u^{k}\ .
\ee
Next, we situate the point of observation on the world line of
charge:
\bd
x^{i} = x^{i}(\tau_{1})\ ;\ u^{k} = u^{k}(\tau_{1})\ .
\ed

Apparently, the electromagnetic field at the point situated on the
trajectory of the particle is created by the particle situated at
this point. Therefore we may expand the integrand in a power series
over $(\tau - \tau_{1})$ . The first, divergent term proportional to
the $du^{i}/ds$ is removed via renormalization of particle mass. The
next term, proportional to the $d^{2}u^{i}/ds^{2}$ is the Dirac -
Lorentz force.  The other terms of the expansion are equal to zero.

Let us apply this procedure to our case.  Electromagnetic field and
self-interaction force are split into two parts in accordance with
decomposition in (9). The first part leads to the Dirac - Lorentz
force.  Indeed, the two points at the infinitely short distance are
connected by the only geodesic line. By $\Delta\varphi \rightarrow 0$
and $\nu < 2$ ( this case corresponds to the real physical situation
[8]) from condition (12) it follows that $n=0.$

The second term, being the gravitational-induced self-interaction
force, is regular on the charge. Therefore, for simplicity we
calculate only the electromagnetic field potential at the location of
particle $x^{i}=x^{i}(\tau_{1})$. They have the following form
\be
A_{2 k}(x^{i}(\tau_{1}))= e\nu\frac{\sin(\pi\nu)}{\pi}
\int_{-\infty}^{+\infty}d\tau\frac{\theta(\Delta t)\theta(q-1)
N_{k}}{\rho(\tau_{1})\rho(\tau)\sqrt{4q(q-1)}}\ ,
\ee
where
\bn
N_{4}&=&u_{4}(\tau)H_{1}(0)\ ,\nonumber \\
N_{z}&=&u_{z}(\tau)H_{1}(0)\ ,\nonumber \\
N_{\rho}&=&u_{\rho}(\tau)H_{1}(1)+u_{\varphi}(\tau)
H_{2}(1)/b\rho (\tau)\ ,\nonumber \\
N_{\varphi}&=&u_{\varphi}(\tau)H_{1}(1)\rho(\tau_{1})/\rho(\tau)
-u_{\rho}(\tau)H_{2}(1)b\rho(\tau_{1})\ ,
\en
and $ H_{1} $ and $ H_{2} $ are determined via the (11)
\be
Q_{\nu}(s)=\frac{\sin(\pi\nu)}{\pi}\{H_{1}(s)-iH_{2}(s)\}\ .
\ee

Let us consider the simple case of the resting particle with the
following trajectory of motion
\be
t(\tau)=\tau\ ,\ z(\tau)=z_{0}\ ,\ \rho(\tau)=\rho_{0}\ ,\
\varphi(\tau)
=\varphi_{0}\ .
\ee
At first, we calculate $A_{2 k}$ at the arbitrary point with
coordinates $(t, z, \rho, \varphi)$. Substituting the trajectory
(17) in (14)-(16) and defining a new variable $y=\ln\beta(\tau)$ we
obtain
\bn
A_{2 4}&=&- e\nu\frac{\sin(\pi\nu)}{\pi}\int_{0}^{+\infty}
\frac{dy}{\sqrt{\Delta z^{2}+\Delta\rho^{2}+4\rho\rho_{0}
\cosh^{2}(y/2)}}\nonumber \\
&\times& \frac{\cosh(\nu y)\cos(\Delta\varphi)-
\cos(\pi\nu)}{[\cosh(\nu y)-\cos(\Delta\varphi+\pi\nu)]
[\cosh(\nu y)-\cos(\Delta\varphi-\pi\nu)]}\ , \\
A_{2 \alpha}&=&0\ .\nonumber
\en
At the location of the charge the field (18) has form
\be
A_{4}=L\frac{e}{\rho_{0}}\ ,\ A_{\alpha}=0\ ,
\ee
\bd
L(\nu)= - \frac{\nu\sin(\pi\nu)}{2\pi}\int_{0}^{+\infty}\frac{dy}
{\cosh(y/2)}\frac{1}{\cosh(\nu y)-\cos(\pi\nu)}\ .
\ed

Thus the gravitational induced self - interaction force has the
only component
\be
{\cal F}_{2}^{\rho} = -{\cal F}_{2 \rho} = -e\partial_{\rho}
A_{2 4} =L
\frac{e^{2}}{\rho_{0}^{2}}\ .
\ee
When $\nu-1$ is small we have the following result :
\be
L \approx \frac{\pi(\nu-1)}{8} \approx \frac{\pi(1-b)}{8} =
\frac{\pi G \mu}{2c^{2}}\ .
\ee

Next we calculate $A_{1 k}$ at the point with coordinates
$t=\tau_{1}, z=z_{0}, \varphi=\varphi_{0}, \rho=\rho_{1}.$ Let us
consider the $\nu < 2 $ . Then from condition (12) we have $n=0$ ,
and
\be
A_{1 4} = \frac{e}{|\rho_{0}-\rho_{1}|}\ ,\ A_{1 \alpha} = 0\ .
\ee

Thus in this approach the self-interaction force is different from
that calculated in Ref.11. The difference is connected with the
coefficient $L$ which may be obtained from results of article [14].

It is possible to calculate self - interaction force for the
arbitrary values of $\nu>1.$ In this case the first part of potential
$A_{1 4}$ for the trajectory (17) contains both non-regular part (22)
and regular part. Eventually the regular on the particle part of
potential has the following form :
\be
A_{4}=L_{0}\frac{e}{\rho_{0}}\ ,\ A_{\alpha}=0\ ,
\ee
where
\be
L_{0}(\nu)=\sum_{n=1}^{[\nu/2]}\int_0^1\frac{dq}{\sqrt{q}}
\delta(q-\sin^2(\pi n/\nu))+L(\nu).
\ee
The coefficient (24) have the simple form when $\nu$ is integer:
\bn
L_{0}(2k+1)&=&\sum_{n=1}^{k}|\sin\frac{\pi n}{2k+1}|^{-1}\ ,
 \nonumber \\
L_{0}(2k)&=&\frac{1}{2}+\sum_{n=1}^{k-1} |\sin\frac{\pi n}{2k}|
^{-1}\ .
\en
 From this expression one can see that the self - interaction force
can build up to high values when the $\nu$ increase. Notes, as
$\nu\to\infty,$ the spacetime of cosmic string tends to the cylindric
spacetime [7].
\section{Remarks and Conclusion}
In this article we investigated the forces acting on the charged
particles in the spacetime of a straight cosmic string.
Electromagnetic potential is split in two parts (9). The first part
is non-regular at the location of the particle and leads to the Dirac
- Lorentz force. The second non-local part is regular on the charge
and leads to the additional gravitational - induced force.
Electromagnetic potential (14) and, consequently, the
self-interaction force, depends on the past history of the charge. In
the simple case of the resting particle this force repels the
particle from the string.

It must be emphasized that the foregoing interaction leads to the
scattering of charged particles by the cosmic string alongside the
scattering of matter from cosmic string of radius $R$ [18] and
Aharonov - Bohm scattering of fermions [19]. By virtue of the fact
that the interaction between charge and string is the Coulomb
repulsion (20) the scattering cross section has the following form
\bd
d\sigma_{st} = L_{0}^{2}d\sigma_{r} = L_{0}^{2}\left(\frac{e^{2}}
{2\varepsilon}\right)^{2}\frac{\cos (\frac{\theta}{2})}{\sin^{3}
(\frac{\theta}{2})} d\theta\ ,
\ed
where $d\sigma_{r}$ is Rutherford cross section, $\theta$ is angle of
scattering, $\varepsilon$ is energy of particle before scattering and
$L_{0}$ is given by the equation (24).\\
\section*{Acknowledgments}
I thank B Linet for helpfull comments on this manuscript. I also
thank the M~L~Newens~B~A for proof reading.
\section*{Appendix}
In order to obtain (9) we must to take into account the following
integrals and series: $2.5.25(9) , 5.4.12(1)$ from [15]
,$2.12.42(16)$ from [17], and the well-known expression in the theory
of distributions
\bd
\sum_{n=-\infty}^{+\infty}e^{inx} = 2\pi\sum_{n=-\infty}^{+\infty}
\delta(x+2\pi n)\ .
\ed

\end{document}